\documentclass[conference]{IEEEtran}
\IEEEoverridecommandlockouts
\usepackage{cite}
\usepackage{amsmath,amssymb,amsfonts}
\usepackage{algorithmic}
\usepackage{graphicx}
\usepackage{textcomp}
\usepackage{xcolor}
\def\BibTeX{{\rm B\kern-.05em{\sc i\kern-.025em b}\kern-.08em
    T\kern-.1667em\lower.7ex\hbox{E}\kern-.125emX}}
\usepackage{tabularx} 
\newcolumntype{Y}{>{\centering\arraybackslash}X} 

\usepackage{multirow} 
        
\usepackage{fancyhdr}
\begin{document}

\title{End-To-End Prediction of Emotion From Heartbeat Data Collected by a Consumer Fitness Tracker}

\author{\IEEEauthorblockN{Ross Harper}
\IEEEauthorblockA{\textit{Limbic} \\
London, UK \\
ross@limbic.ai}
\and
\IEEEauthorblockN{Joshua Southern}
\IEEEauthorblockA{\textit{Limbic} \\
London, UK \\
josh@limbic.ai}

}

\maketitle
\thispagestyle{fancy}

\begin{abstract}
Automatic detection of emotion has the potential to revolutionize mental health and wellbeing. Recent work has been successful in predicting affect from unimodal electrocardiogram (ECG) data. However, to be immediately relevant for real-world applications, physiology-based emotion detection must make use of ubiquitous photoplethysmogram (PPG) data collected by affordable consumer fitness trackers. Additionally, applications of emotion detection in healthcare settings will require some measure of uncertainty over model predictions. We present here a Bayesian deep learning model for end-to-end classification of emotional valence, using only the unimodal heartbeat time series collected by a consumer fitness tracker (Garmin V\'ivosmart 3). We collected a new dataset for this task, and report a peak F1 score of 0.7. This demonstrates a practical relevance of physiology-based emotion detection `in the wild' today.
\end{abstract}


\section{Introduction}
Analysis of human emotion is the bedrock of affective computing. The majority of research in this field has focussed on predicting emotion from face, voice and text \cite{Pantic2000, Hanjalic2005, ElKaliouby2005, Yang2008, Zeng2009, Schuller2010, Polzehl2011, Schuller2011}. Physiological analysis has garnered comparatively little attention \cite{Kim2008b,Alzoubi2012, Goshvarpour2017}, and explores the neurobiological correlates of emotion within the limbic and autonomic nervous systems. 

Physiology-based emotion detection has tremendous potential to compliment existing methods of affective computation. For instance, analysis of face, voice, and text rely heavily on expression, which can vary across individuals and cultures \cite{Ekman1987, Scherer2001}, and can also be easily faked. By comparison, physiological processes are far less volitional. Physiological analyses present a further opportunity for non-invasive continuous monitoring - as physiological signals may be passively measured throughout the day. 

For these reasons, physiology-based emotion detection has the capacity to fill critical gaps in domains where it is challenging to continuously collect audiovisual data (e.g. healthcare). Perhaps unsurprisingly, there is a growing physiological resurgence within affective computing. The vast majority of studies rely on a combination of autonomic markers to classify emotional response. These include galvanic skin response (GSR), electroencephalogram (EEG), electromyogram, respiration, skin temperature (ST) and electrocardiogram (ECG). While attractive from a modelling perspective, such multimodal input may not always be available. A select few studies have therefore constructed models capable of predicting emotion from \textit{unimodal} ECG data \cite{harper2019bayesian, Katsigiannis2018, Subramanian2016, Keren2017, Miranda-Correa2017b, Guo2016a, Ferdinando2016, Valenza2014b, Agrafioti2012}. The idea here tends to be that such models could be extended to predict emotion using wearable heart monitoring devices `in the wild'. Indeed, it has been shown that emotional valence can be classified using expensive lab-based wearable ECG recording devices \cite{harper2019bayesian}. However, to be truly relevant for large-scale real-world monitoring today, such classifiers must be compatible with the growing number of affordable consumer fitness trackers. These almost exclusively extract heartbeat from photoplethysmogram (PPG). 

Consumer fitness trackers typically extract the peaks of the PPG signal to obtain a heartbeat time series, or `inter-beat intervals' (IBIs). As IBIs can be extracted from ECG and PPG data, we use the notation IBI\textsubscript{ECG} and IBI\textsubscript{PPG} to distinguish between the two. To the best of our knowledge, no previous work has rigorously explored the suitability of IBI\textsubscript{PPG} generated by affordable fitness trackers for predicting human emotion at scale. Such research has the potential for immediate real-world application, as the number of wrist-worn wearable devices continues to rise into the hundreds of millions   \cite{idcmedia} and all major brands now incorporate commoditised PPG sensors as standard (e.g. Fitbit, Polar, Samsung Gear, Apple Watch, and Garmin).

In this study, we use a Bayesian deep neural network model that was shown previously to classify emotional valence from IBI\textsubscript{ECG} \cite{harper2019bayesian}. We extend this model for emotion detection using IBI\textsubscript{PPG} collected by a consumer wearable. For this study, we generated a new dataset comprising IBI\textsubscript{PPG} data (collected using a Garmin V\'ivosmart 3 device). This data was recorded during presentation of a number of short emotion-inducing video stimuli. We explore the statistical differences between this IBI\textsubscript{PPG} and previously collected IBI\textsubscript{ECG}. We go on to show that training a neural network classifier on IBI\textsubscript{ECG} confers no performance improvement when tested on IBI\textsubscript{PPG}, demonstrating the necessity for new datasets built around cheap off-the-shelf wearable devices.

\section{Related Work}
This section provides an overview of relevant work, with a focus on (A) unimodal ECG for emotion prediction, and (B) unimodal PPG for emotion prediction. 

\subsection{Emotion Prediction from Unimodal ECG Data} 
Existing approaches for prediction of emotion using physiological signals typically pool a number of bio-signals to provide multimodal input to a classifier algorithm \cite{Kim2008b, Alzoubi2012, Goshvarpour2017}. Fewer studies narrow their scope to unimodal ECG input in accordance with the heartbeat-centric limitations of affordable wearable devices. Additionally, those studies that have explored unimodal heartbeat models for emotion detection tend to ignore temporal structures of the signal. Instead, they use `static' classification methods that analyse global features of the input time-series, such as Naive Bayes (NB), \cite{Miranda-Correa2017b, Subramanian2016}, linear discrimant analysis (LDA) \cite{Agrafioti2012}, and support vector machine (SVM) \cite{Katsigiannis2018, Guo2016a, Valenza2014b}. A summary of these studies can be found in Table \ref{relevant_work}. Two notable exceptions have implemented temporal neural network models to predict emotional valence from ECG input \cite{Keren2017, harper2019bayesian}. In these examples, convolutional and recurrent network layers were used to perform end-to-end learning, which improved upon computationally expensive manual feature engineering schemes. In \cite{harper2019bayesian}, a Bayesian framework was further used to output probability distributions over valence predictions, making this model particularly suited for applications in domains such as healthcare, where a high premium is placed on predictive certainty.

\begin{table*}[!t]
	\renewcommand{\arraystretch}{1.3}
	\caption{Summary of relevant work}
	\label{relevant_work}
	\centering
	\begin{tabularx}{\linewidth}{|Y|Y|c|Y|Y|Y|}
		\hline
		\textbf{Author} & \textbf{Stimulus} & \textbf{Subjects} & \textbf{Model} & \textbf{Target} & \textbf{Performance}\\
		\hline
		Harper \& Southern 2018 \cite{harper2019bayesian} & Videos & 40 & LSTM and CNN & High/Low Valence & Acc. 90\% (Chance: 50\%) \\
		\hline
		Katsigiannis \& Ramzan 2018 \cite{Katsigiannis2018} & Videos & 23 & SVM & High/Low Valence & F1. 0.5305 (Chance: 0.500) \\
		\hline
		Subramanian et al 2018 \cite{Subramanian2016} & Videos & 58 & NB & High/Low Valence & Acc. 60\% (Chance: 50\%) \\
		\hline
		Keren et al 2017 \cite{Keren2017} & Naturalistic dyadic interactions & 27 & LSTM and CNN & Continuous Valence (Regression) & Concordance Correlation Coefficient. 0.210 (Baseline: 0.121)\\
		\hline
		Miranda-Correa et al 2017 \cite{Miranda-Correa2017b} & Videos & 40 & NB & High/Low Valence & F1. 0.545 (Chance: 0.500) \\
		\hline
		Guo et al 2016 \cite{Guo2016a} & Videos & 25 & SVM & High/Low Valence & Acc. 71.40\% (Chance: 50\%) \\
		\hline
		Ferdinando et al 2016 \cite{Ferdinando2016} & Videos \& Images & 27 & KNN & High/Medium/Low Valence & Acc. 59.2\% (Chance: 33.3\%) \\
		\hline
		Valenza et al 2014 \cite{Valenza2014b} & Images & 30 & SVM & High/Low Valence & Acc. 79.15\% (Chance: 50\%) \\
		\hline
		Agrafioti et al 2012 \cite{Agrafioti2012} & Images & 32 & LDA & Gore, Erotica & Acc. 46.56\% (Chance: 50\%) \\
		\hline
	\end{tabularx}
\end{table*}

\subsection{Emotion Prediction from Unimodal PPG Data} 
Very few studies have explored emotion detection with a focus on PPG data. One study combined GSR and PPG, collected by a Shimmer3 sensor \cite{shimmer:xxx}, to classify High/Low valence and arousal \cite{Udovicic:2017:WER:3132635.3132641}. In another study, unimodal PPG data collected by an expensive wrist-worn wearable device (Empatica E4 \cite{7015904}) was compared with data collected by a laboratory sensor (Biopac MP150 \cite{biopac:xxx}) \cite{10.1007/978-3-319-60639-2_2}. Although these studies represent an important step towards real-world applicability, we are not aware of any studies that have explored emotion recognition using IBI\textsubscript{PPG} data of the type collected by affordable consumer fitness trackers.

\section{Experimental Setup}
In this section, we describe the experimental procedure for collecting IBI\textsubscript{PPG} from a consumer fitness tracker (Garmin V\'ivosmart 3).

\subsection{Experimental Protocol}
We used an emotion-inducing stimulus setup combined with participant self-reporting, as is conventional within the field of affective computing. The experiment involved 17 study participants (5 female; 12 male). Each participant received an initial tutorial on how to self-report their emotional state using the widely-used Self-Assessment Manikin (SAM) framework for measuring emotion \cite{Morris1995}. Next, the Garmin V\'ivosmart 3 was secured to the left wrist of the participant, and IBIs extracted by the embedded PPG sensor were collected. The participants were seated directly in front of a computer screen (at a distance of 60cm) and were asked to wear headphones in order to reduce external distractions. The experimenter then left the room and the recording session began.

\begin{figure}\centering
	\includegraphics[width=\linewidth]{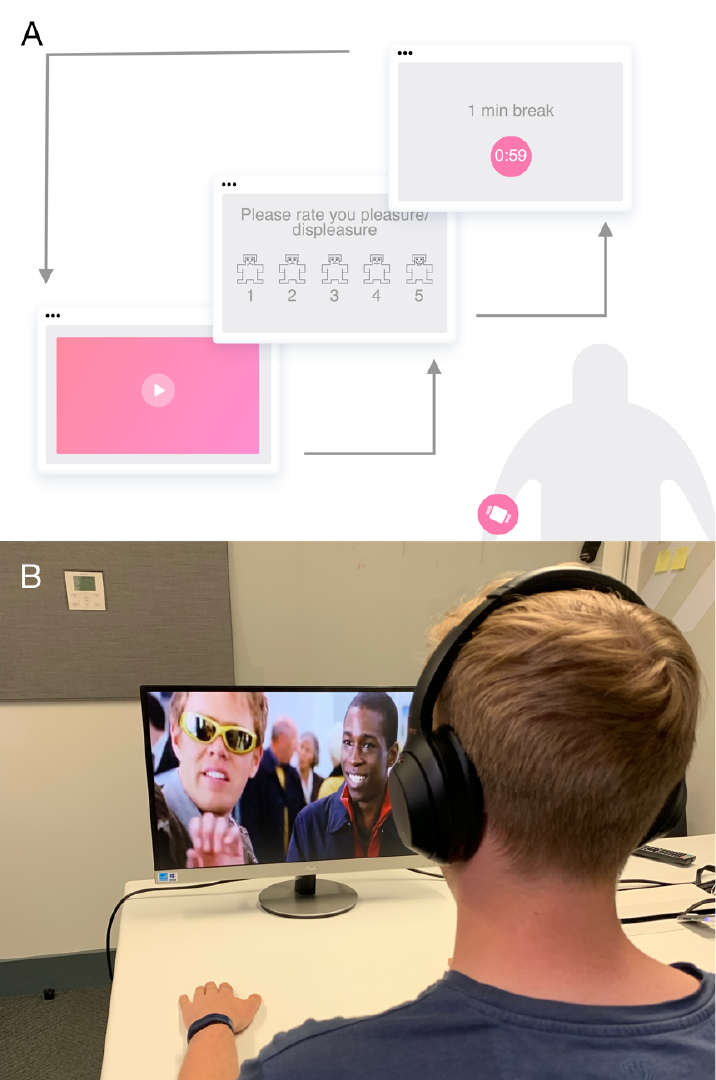}
	\caption{Experimental setup. The participant was seated in front of a single computer monitor. A video stimulus from Table \ref{tab:Videos} was selected randomly and played. After the stimulus had ended, the participant was asked to report their emotional valence using the SAM framework. The next randomly selected video stimulus was presented after a one minute break. This process repeated until all 24 stimuli had been presented.}
	\label{fig:Experimental_Setup}
\end{figure}

Emotion-inducing video stimuli were presented on the computer screen in randomised order. At the end of each video stimulus, the participant was asked to complete an emotional valence self-report using the SAM framework \cite{Morris1995}. After completing the emotion self-report, the participant experienced one minute of a neutral scene and was asked to clear their mind as much as possible prior to the next video stimulus. This was done to reduce carry-over of emotions between video stimuli. A schematic overview of the experimental setup can be found in Fig.~\ref{fig:Experimental_Setup}A, with photograph shown in Fig.~\ref{fig:Experimental_Setup}B.

\subsection{Stimuli} 
\label{section:Video_Stimuli}
The participants each viewed 24 short video stimuli presented in random order. 96 potential videos were initially chosen, and independently annotated for emotional valence by 30 volunteers using the SAM framework. The variance of these annotations was then calculated for each video, and the 96 potential videos ranked from lowest to highest variance (lowest variance at the top, representing highest agreement amongst the 30 annotators). The top 25\% of videos were selected (24 stimuli). Of these, 8 videos had been independently scored as inducing pleasure (high valence); 16 videos were independently scored as inducing displeasure (low valence). The average stimulus length was 02:29 (See Table \ref{tab:Videos}). 

\begin{figure*}[ht]\centering
	\includegraphics[width=\linewidth]{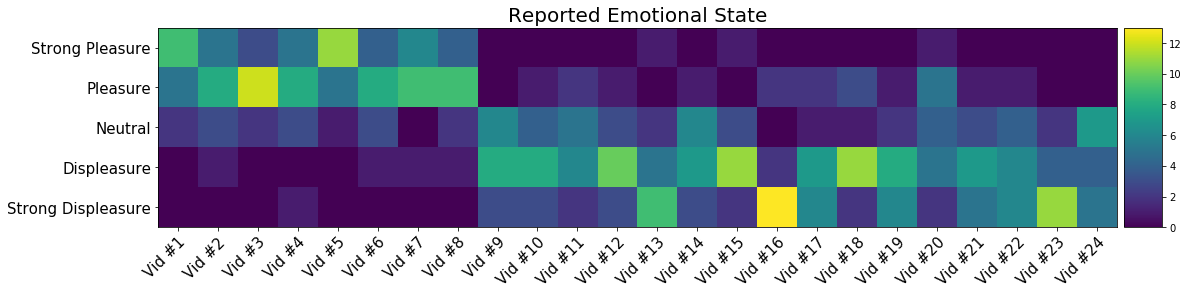}
	\caption{Self-reported emotional valence induced by each video clip. Study participants rated their emotional state after each video clip on a five-point scale from `Strong Displeasure' to `Strong Pleasure' in accordance with Self-Assessment Manikin (SAM) framework for measuring emotion \cite{Morris1995}. The image shows the density of reports for each emotional state: blue to yellow with greater density of ratings (see colour bar). Note the videos 1-8 elicited more pleasurable emotions compared to 9-24, which elicited more displeasurable emotions. This is in agreement with Table \ref{tab:Videos}.}
	\label{fig:Reported_Emotional_State}
\end{figure*}

To confirm that the 24 test videos induced the expected emotional valence in study participants, we show in Fig.~\ref{fig:Reported_Emotional_State} the density of valence scores obtained from study participants during the experiment. We see that these self-reports broadly match those of the 30 volunteer annotators (Table \ref{tab:Videos}). All video stimuli were presented, and the SAM administered, using a custom-made web app.  

\subsection{Measuring PPG} \label{section:Measuring_PPG}
IBI data extracted from the PPG signal was collected using the Garmin V\'ivosmart 3, which retails at around £70 (\$90). For this, we developed a custom Android Wear app using the Android Wear SDK. This app collected the IBIs locally on a mobile device, synchronised these with the timing of video stimuli presentation, and sent the resulting data files to cloud servers upon experiment completion.

\subsection{Facial Video Recordings}
Frontal face video was also recorded during the experiment using a web-cam positioned centrally on the computer screen. Although our present study does not incorporate visual data for affect recognition, this data can be used for future work comparing facial and physiological signals for prediction of emotion.

\begin{table}[h]
	\caption{Selected Video Stimuli}
	\label{tab:Videos}
	\begin{tabularx}{\linewidth}{|c|Y|Y|}
		\hline
		\textbf{Video} & \textbf{Elicitation} & \textbf{Duration} \\ 
		\hline
		\#1 & Pleasure & 0:53 \\
		\#2 & Pleasure & 3:52 \\
		\#3 & Pleasure & 1:14 \\
		\#4 & Pleasure & 2:15 \\
		\#5 & Pleasure & 2:39 \\
		\#6 & Pleasure & 4:28 \\
		\#7 & Pleasure & 0:51 \\
		\#8 & Pleasure & 1:11 \\
		\hline
		\#9 & Displeasure & 2:11 \\
		\#10 & Displeasure & 6:08 \\
		\#11 & Displeasure & 1:22 \\
		\#12 & Displeasure & 2:06 \\
		\#13 & Displeasure & 1:51 \\
		\#14 & Displeasure & 1:47 \\
		\#15 & Displeasure & 2:53 \\
		\#16 & Displeasure & 5:09 \\
		\#17 & Displeasure & 0:56 \\
		\#18 & Displeasure & 2:13 \\
		\#19 & Displeasure & 2:47 \\
		\#20 & Displeasure & 4:52 \\
		\#21 & Displeasure & 0:40 \\
		\#22 & Displeasure & 2:19 \\
		\#23 & Displeasure & 2:51 \\
		\#24 & Displeasure & 2:18 \\ \hline	
	\end{tabularx}
\end{table}

\section{Model} \label{section:Model}

\subsection{Neural Network Architecture}
Deep neural networks have obtained promising results for end-to-end classification of valence from unimodal ECG data \cite{Keren2017, harper2019bayesian}. In this study, we use the neural network architecture described in \cite{harper2019bayesian}, which incorporates a Bayesian framework to model probability distributions over model output. (For details of the model hyperparameters and training protocol, please see the original text). 

\begin{figure}
	\centering
	\includegraphics[width=\linewidth]{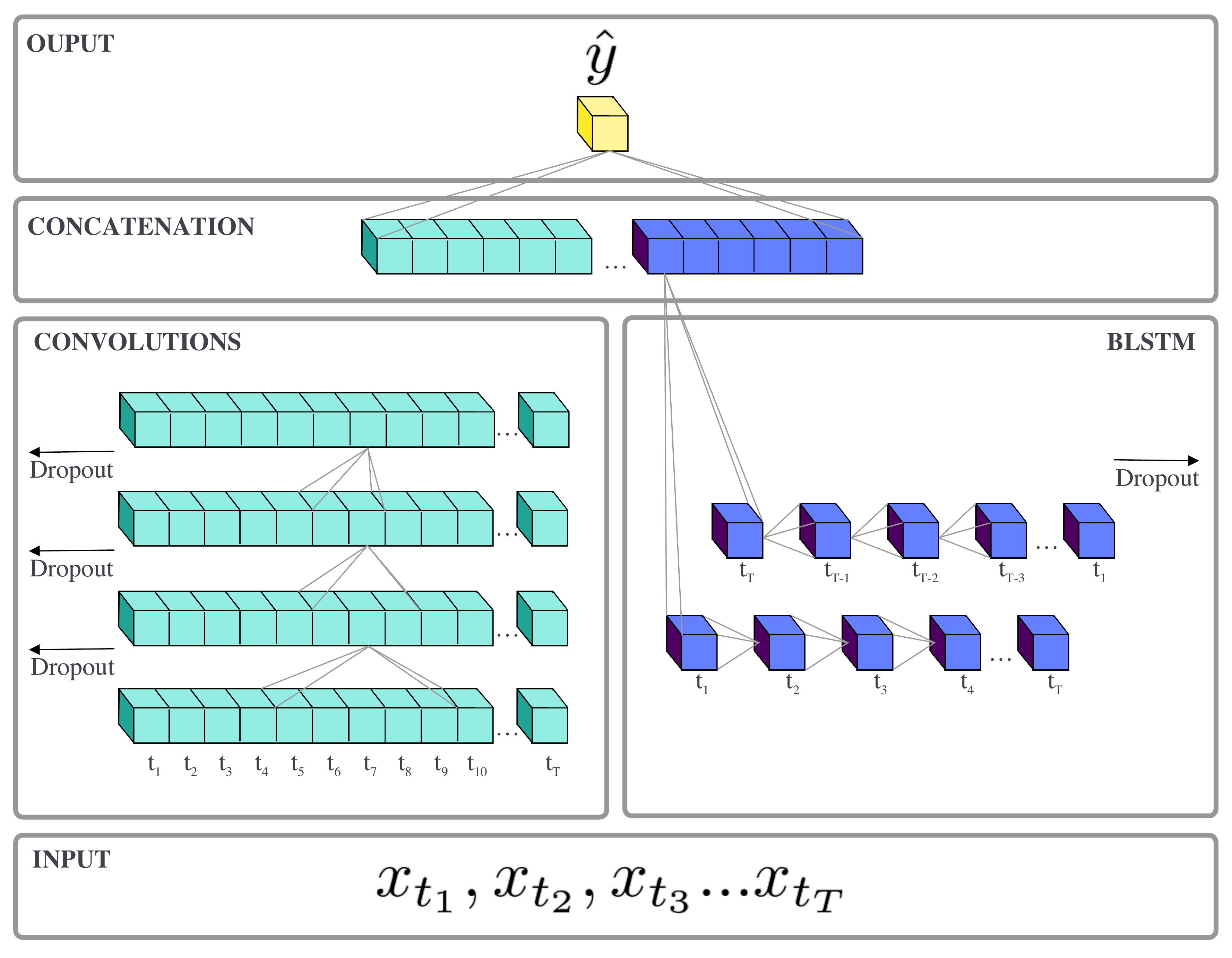}
	\caption{End-to-end model architecture (adapted from \cite{harper2019bayesian}). Data flows through two temporal processing streams: 1D convolutions (green) and a bi-directional LSTM (blue). The output from both streams is then concatenated before passing through a dense layer to output a regression estimate for valence, $\hat{y}$. }
	\label{fig:ModelArchitecture}
\end{figure}

An overview of our model is shown in Fig.~\ref{fig:ModelArchitecture}. In brief, the IBI time series passes through two concurrent streams. The first stream comprises four stacked convolutional layers with filter size set to 128, and window size decreasing from 8 to 2 time steps with network depth. This extracts features from larger receptive fields as the data passes through each successive layer. Monte Carlo dropout is applied after each convolutional layer, as well as a ReLU activation function (for details, see \cite{harper2019bayesian}). 

The second stream comprises a bidirectional LSTM (each with 32 hidden units), also followed by Monte Carlo dropout. This recurrent structure permits temporal modelling of the heartbeat time series, which is non-linear and non-stationary \cite{Weber1992, Sunagawa1998}. The output of these two streams is finally concatenated into a 192-length vector (128 from the convolutional stream; 64 from the LSTM stream) before passing through a dense layer to output a regression estimate for valence.

Uncertainty is a key component of decision-making in many real-world domains, especially healthcare \cite{Ghahramani2015}. It therefore follows that applications of physiology-based emotion detection in this area must incorporate probabilistic considerations. We therefore use Monte Carlo dropout to recast our neural network as a Bayesian model, performing $N$ stochastic forward passes through the network to approximate a posterior distribution over model predictions \cite{Gal2016}.

\subsection{Binary Classification Framework}
In order to translate from a regression to a classification scheme, we introduce decision boundaries in continuous space. For a binary (high/low) classification, this can be done by including a decision boundary at the central point of the valence axis. We next introduce a confidence threshold parameter, $\alpha$, to tune predictions to a specified level of model uncertainty. For example, when $\alpha = 0.95$, at least $95\%$ of the output distribution must lie above or below the valence scale midpoint in order for the input sample to be classified as belonging to the high or low valence class respectively. If this is not the case, no prediction is made (the model respectfully makes no comment). As our model may not classify all instances, we adopt the term ‘coverage’ to denote the set of cases for which it is confident enough to make a prediction. For an in-depth discussion, see \cite{harper2019bayesian}.

Note that for a binary classification problem, and $N$ is an odd integer, there will always be at least $50\%$ of the output distribution above or below the valence midpoint. Thus, when $\alpha = 0.5$, classification is determined by the median of the output distribution, and the coverage is $100\%$. As $\alpha$ increases, model behaviour moves from risky to cautious $-$ lower coverage, but more confidence in the classification. This aligns with our goal of providing real-world relevance to physiology-based emotion prediction in domains such as healthcare.

\section{External Data}
We applied the Bayesian deep learning framework described above (and in \cite{harper2019bayesian}) to achieve end-to-end prediction of emotion using IBI\textsubscript{PPG} collected by the Garmin V\'ivosmart 3. However, we further wished to explore the differences between these IBI\textsubscript{PPG} and IBI\textsubscript{ECG} extracted from a laboratory-grade monitor. For this comparison, we used the established AMIGOS dataset \cite{Miranda-Correa2017b}.

The AMIGOS dataset consists of 40 healthy participants (13 female; 27 male) aged between 21 and 40 years old (mean: 28.3). The ECG was recorded using a Shimmer\textsuperscript{TM} ECG wireless monitoring device (256 Hz, 12 bit resolution) \cite{shimmer:xxx}. The participants watched 18 film clips (duration $< 395$ seconds), which had been selected for their ability to elicit strong emotional responses \cite{Miranda-Correa2017b}. The videos were presented to the subjects in a random order with a 5-second baseline recording of a fixation cross being shown before each video. Each film clip was followed by self-assessment of valence on a scale of 1 to 9 using SAM \cite{Morris1995}.  

\section{Methods}	
\subsection{Pre-processing}	
The IBI\textsubscript{PPG} extracted from the PPG sensor in the Garmin V\'ivosmart 3 were z-score normalized and zero padded to the length of the longest training sample. For the AMIGOS data, IBIs were extracted manually from the ECG time-series using a combined adaptive threshold method \cite{Christov2004}. The resulting IBI\textsubscript{ECG} was then also z-normalized and zero padded or cut to the length of the longest IBI\textsubscript{PPG} training sample. 

\subsection{Training and Hyperparameters} \label{section:Training_and_Hyperparams}
The hyper-parameters of the model were set to those specified previously \cite{harper2019bayesian}. The convolutional kernels were initialized as He normal \cite{He2015} with a filter size set to 128, and a window size decreasing from 8 to 2 time steps with network depth. A dropout of $50\%$ was applied after each convolutional block, and $80\%$ dropout followed the bi-directional LSTM, which comprised 32 hidden units. The training phase was run for 1000 epochs using Adam optimization \cite{Kingma2014} and the learning rate decreased from $e^{-3}$ to $e^{-4}$, halving with a patience of 100 epochs. The model was implemented using Tensorflow \cite{GoogleResearch2015}.      

\subsection{Evaluation}
Model performance was assessed using 10 iterations of leave-one-subject-out cross-validation to show the ability of the model to generalize to new people. For each iteration, one subject was randomly selected and their data held out as a test set. Dropout was applied at test time with $N = 1000$ forward propagations made through the network to generate an empirical distribution over model output. As outlined in section \ref{section:Model}, a given test input sample was classified into a binary high/low valence class provided a proportion of at least $\alpha$ posterior distribution mass fell above or below the valence midpoint respectively. If this was not the case, then no prediction was made. The model's F1 score was then calculated based on those classifications that the model attempted. We chose to evaluate our model using the F1 score, rather than accuracy, due to the unbalanced high/low valence videos in the dataset (selected as described in Section \ref{section:Video_Stimuli}).  

\section{Results}
\subsection{Comparison of IBIs Extracted from ECG and PPG}
In order to gain an understanding of the differences between IBIs extracted from ECG data (collected by the commonly-used laboratory-grade Shimmer\textsuperscript{TM}), and IBIs extracted from a consumer PPG sensor (Garmin V\'ivosmart 3), we calculated a number of features for all IBI samples across both datasets. 

Frequency domain features included (1) spectral power in the frequency range [0.15, 0.4] Hz (HF power), (2) spectral power in the frequency range [0.04, 0.15] Hz (LF power), spectral power in the frequency range [0.003, 0.04] Hz (VLF power), and (4) ratio of low frequency to high frequency signal (LF/HF). Time domain features included (1) mean, (2) median, (3) standard deviation (SDSD), (4) number of instances where the change between successive IBIs is greater than 0.02 (NN20), (5) normalised NN20 (pNN20), (6) the root mean square of the successive differences (rMSSD), and (7) the multiscale entropy. 

\begin{figure}
	\centering
	\includegraphics[width=\linewidth]{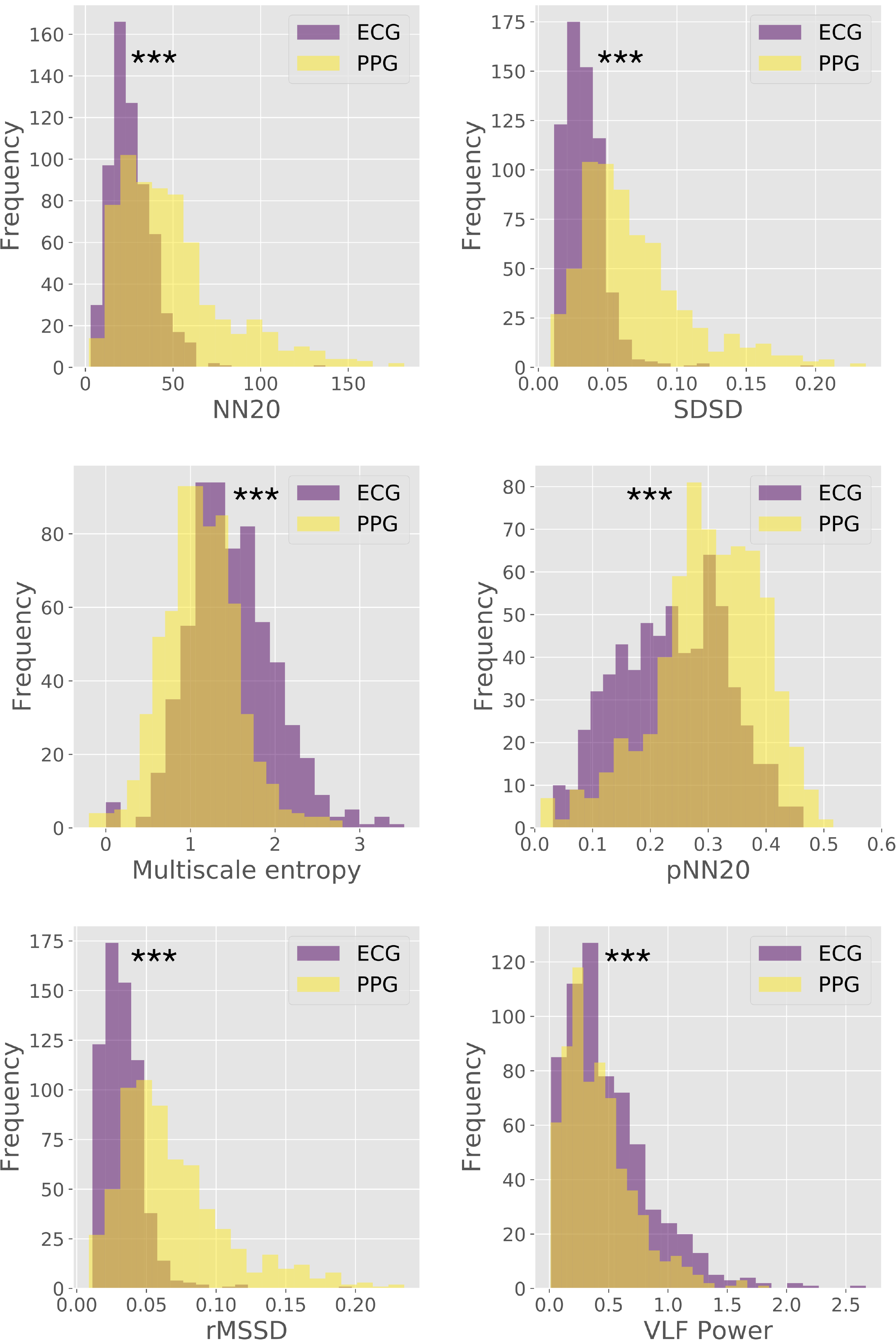}
	\caption{Histograms showing the values of different features calculated from the time-series of both the IBI\textsubscript{PPG} (yellow) and IBI\textsubscript{ECG} data (purple).}
	\label{fig:StatisticalDifferences}
\end{figure}

Non-parametric Mann-Whitney test was performed for each feature to identify statistically significant differences between IBIs extracted from ECG and PPG. Statistically significant differences were observed for VLF power, SDSD, NN20, pNN20, rMSSD and the multiscale entropy (See Fig.~\ref{fig:StatisticalDifferences} and Table \ref{tab:pvalues}). 

To further probe these statistical differences, a simple SVM classifier was used to differentiate IBIs extracted from ECG and PPG using the previously calculated features as input. The sklearn library in Python \cite{Pedregosa:2011:SML:1953048.2078195} was used to build a C-Support Vector Classification with `rbf' kernel and penalty parameter, C, set to 1. 10-fold cross-validation was implemented and accuracy of the classifier was found to be $70\%$. This supports the conclusion that there are structural differences in the statistical properties between IBI\textsubscript{PPG} and IBI\textsubscript{ECG}.

\boldmath
\begin{table}[h]
	\caption{Mann-Whitney P-values between IBI\textsubscript{ECG}  and IBI\textsubscript{PPG} features}
	\label{tab:pvalues}
	\begin{tabularx}{\linewidth}{|c|Y|Y|}
		\hline
		& \textbf{Feature} &   \textbf{P-value} \\ \hline
		1 & HF power & 0.22 \\ \hline
		2 & LF power & 0.22 \\ \hline
		3 & \textbf{VLF power} & $< 0.001$ \\ \hline
		4 & LF/HF & 0.22 \\ \hline	
		6 & Mean & 0.27 \\ \hline
		7 & Median & 0.26 \\ \hline
		8 & \textbf{SDSD} & $< 0.001$ \\ \hline
		9 & \textbf{NN20} & $< 0.001$ \\ \hline
		10 & \textbf{pNN20} & $< 0.001$ \\ \hline
		11 & \textbf{rMSSD} & $< 0.001$ \\ \hline
		12 & \textbf{Multiscale Entropy} & $< 0.001$ \\ \hline
	\end{tabularx}
\end{table}
\unboldmath 

\subsection{Predicting Emotion Using IBIs Extracted from PPG}
\begin{figure}
	\centering
	\includegraphics[width=\linewidth]{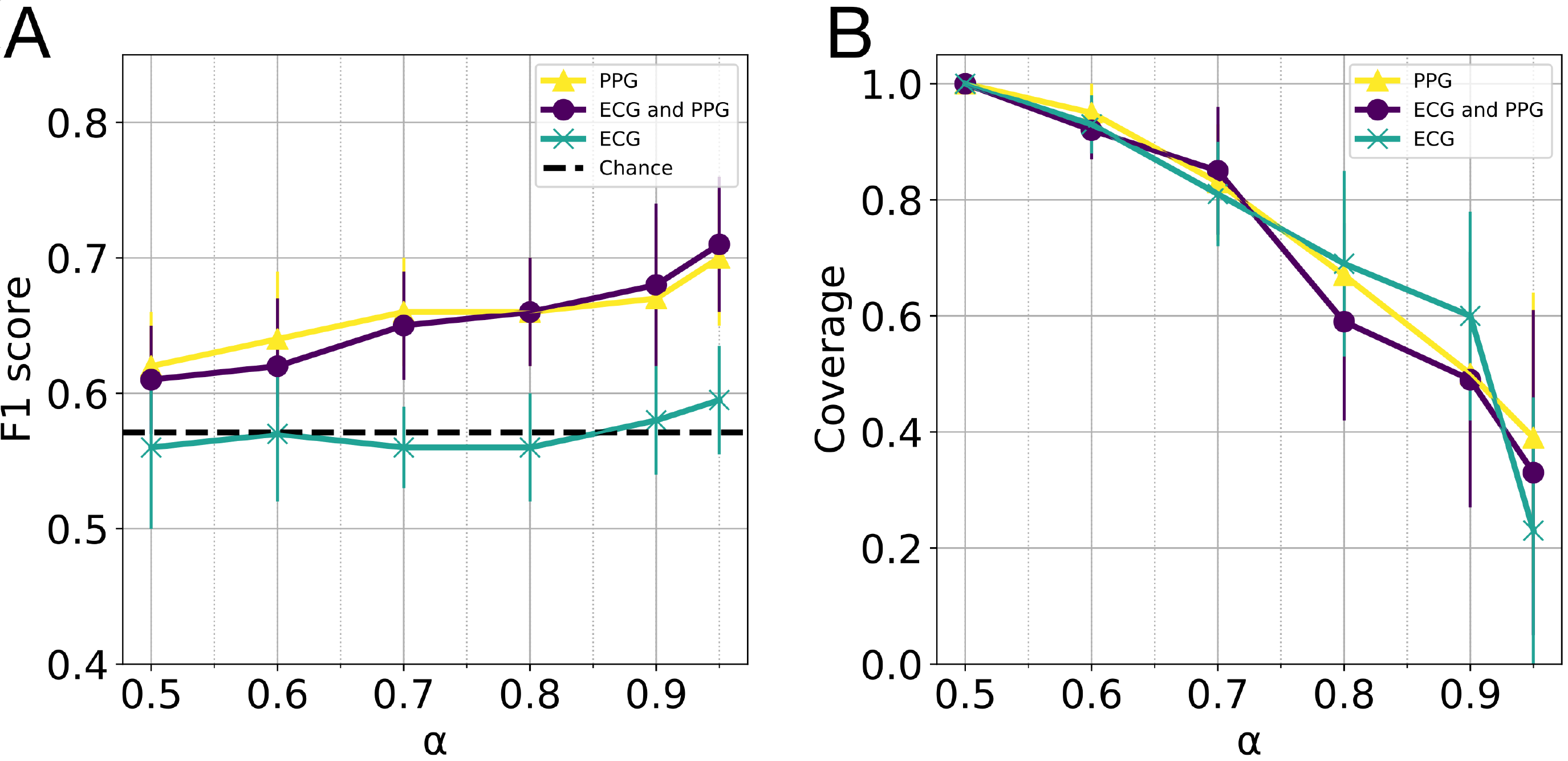}
	\caption{Classification performance of high/low valence using IBI\textsubscript{PPG} input from a consumer wearable. (A) Model F1 score as a function of $\alpha$. (B) Model coverage as a function of $\alpha$. Results are shown for model trained on the IBI\textsubscript{PPG} data alone (yellow, triangles), IBI\textsubscript{ECG} and IBI\textsubscript{PPG} data together (blue, circles), and IBI\textsubscript{ECG} data alone (green, crosses). Grey dashed line in (A) shows F1 score of a chance model for comparison.}
	\label{fig:ModelResults}
\end{figure}

We implemented the Bayesian neural network described in Section \ref{section:Model} using IBI\textsubscript{PPG} data collected by the Garmin V\'ivosmart 3 (see Section \ref{section:Measuring_PPG}). As $\alpha$ increased, so too did the F1 score, demonstrating a clear relationship between model confidence and propensity to make accurate predictions  (Fig.~\ref{fig:ModelResults}A). As expected, model coverage decreased as $\alpha$ increased, due to the fact that fewer output distributions met the necessary threshold for a prediction to be made  (Fig.~\ref{fig:ModelResults}B). When $\alpha = 0.95$, our model achieved a peak F1 score of 0.7 (Fig.~\ref{fig:ModelResults}A).  

\subsection{Further Training with IBIs Extracted from ECG}
We next investigated whether IBI\textsubscript{ECG} collected by the commonly-used laboratory-grade Shimmer\textsuperscript{TM} conferred any advantage to the task of predicting emotion from IBI\textsubscript{PPG} collected by the consumer fitness tracker. The IBI\textsubscript{ECG} data from the AMIGOS dataset was added to the IBI\textsubscript{PPG} training set, and the model was evaluated, as before, on the IBI\textsubscript{PPG} test set (see Section \ref{section:Training_and_Hyperparams} for train-test subdivision). No significant difference was observed in model performance when trained on IBI\textsubscript{PPG} data alone versus IBI\textsubscript{PPG} combined with IBI\textsubscript{ECG} (p = 0.16, computed using Mann-Whitney test between the 10 F1 scores, $\alpha = 0.5$). 

For completeness, we further trained the model using the IBI\textsubscript{ECG} data alone, and then evaluated on the IBI\textsubscript{PPG} test data. In this setting, the model performed no better than chance. (Here, the chance F1 score of 0.57 is the F1 score obtained when a video is naively classified as either high or low valence with equal probability). The performance of the model with these different combinations of training data is shown in Fig.~\ref{fig:ModelResults}. 

\section{Discussion}
The growing prevalence of affordable consumer wearable monitoring devices has created an opportunity for emotion detection at scale. Recent work has tried to bridge the gap from laboratory to real-world through the analysis of unimodal heartbeat data (in accordance with the availability of heartbeat sensors). However, no study has explored affect recognition on heartbeat data collected by a cheap off-the-shelf consumer wearable device. This is important if physiology-based emotion detection is to have immediate relevance today.

In this study, we have shown that the IBI data collected by a popular fitness tracker is statistically different to that which is collected by a widely-used laboratory-grade ECG monitor. Of particular note is that significant differences were found for more time domain features of the heartbeat signal, as compared to frequency domain features. Additionally, the IBI\textsubscript{ECG} data did not confer any performance advantage when used to train our neural network model for the task of predicting valence from IBI\textsubscript{PPG} samples generated by the consumer fitness tracker. This supports the conclusion that real-world applications of physiology-based emotion detection would benefit from new datasets built around cheap off-the-shelf wearable devices. This study represents a good first attempt, which, using a Bayesian neural network classifier, achieved a promising peak F1 score of 0.70 from our new dataset comprising of 17 participants. 

Our probabilistic classification framework includes a confidence parameter, $\alpha$, which allowed the F1 score and coverage of our model to be tuned according to varying demands on prediction certainty. The use of a regression output further allows the experimenter to switch easily between regression and classification tasks, and indeed allows her to specify bespoke decision boundaries appropriate for binary- or multi-class tasks. We chose to incorporate these Bayesian considerations to align with our overarching goal of making physiology-based emotion detection relevant to real-world applications. For instance, emotion detection for mental health monitoring might reasonably require high levels of certainty to predict the onset of major depressive disorder. Additionally, clinical triaging is possible, where uncertain model predictions are sent to a human expert for review (or perhaps a more computationally expensive model). Similar levels of certainty may not, however, be absolutely necessary in many consumer products.

\bibliographystyle{ieeetr}
\bibliography{Bibliography}

\end{document}